\documentclass[aip,jap,reprint,floatfix,groupedaddress]{revtex4-1}
\usepackage{graphicx}
\usepackage{amsmath,amssymb}
\usepackage{hyperref}

\begin{document}

\title{Transparent electrodes with nanorings: a computational point of view} %Title of paper
\author{Mohammad-Reza Azani}
\email{moha@intercomet.com}
\author{Azin Hassanpour}
\email{azin@intercomet.com}
\affiliation{Department of Research and Development, Intercomet S.L.
Calle Ca\~{n}ada, 15, 28860 Paracuellos de Jarama, Madrid, Spain}

\author{Yuri~Yu.~Tarasevich}
\email[Corresponding author: ]{tarasevich@asu.edu.ru}

\author{Irina~V.~Vodolazskaya}
\email{vodolazskaya\_agu@mail.ru}

\author{Andrei~V.~Eserkepov}
\email{dantealigjery49@gmail.com}

\affiliation{Laboratory of Mathematical Modeling, Astrakhan State University, Astrakhan, 414056, Russia}

\date{\today}

\begin{abstract}
Four samples of transparent conductive films with different numbers of silver nanorings per unit area were produced. The sheet resistance, transparency, and haze were measured for each sample.  Using Monte Carlo simulation, we studied the electrical conductivity of random resistor networks produced by the random deposition of the conducting rings onto the substrate. Both systems of equal-sized rings, and systems with rings of different sizes were simulated. Our simulations demonstrated the linear dependence of the electrical conductivity on the number of rings per unit area. Size dispersity decreased the percolation threshold, but without having any other significant effect on the behavior of the electrical conductance. Analytical estimations obtained for dense systems of equal-sized conductive rings were consistent with the simulations.
\end{abstract}


\maketitle %\maketitle must follow title, authors, abstract and \pacs

\section{Introduction}\label{sec:intro}
Transparent electrodes are used in a range of different electronic devices such as touch screens, displays, and solar cells. Recent advances in nano-materials research have offered new, transparent conductive materials, e.g., carbon nanotubes (CNTs), graphene, metal nanowires (NWs), and printable metal grids.\cite{Hecht2011AM} In the case of CNTs and metal NWs, the concentration of conducting object should be large enough to ensure the occurrence of a conducting network connecting opposite borders of the film, i.e., the system has to be above the percolation threshold. Simultaneously, the concentration should be small enough to ensure high transparency of the film. These contradictory requirements have stimulated new research. To characterize both the sheet resistance, $R_\Box$, and the transparency, $T$, the figure of merit (FoM) is used\cite{Haacke1976JAP}
\begin{equation}\label{eq:FoM}
  \Phi_\text{TC} = \frac{T^{10}}{R_\Box}.
\end{equation}
Numerous works have been devoted to the electrical and optical properties of transparent films with elongated fillers such as NTs, NWs, and nanorods.\cite{Yi2004JAP,De2010ACSN,Heitz2011N,Mutiso2013ACSN,Khanarian2013JAP,Khanarian2013JAP,Large2016SR,Callaghan2016PCCP,Kumar2016JAP,Kumar2017JAP,Kim2018JAP,Ainsworth2018ATS,Forro2018ACSN}
However, the use of films containing conducting nanorings\cite{Layani2009ACSN,Shimoni2014,Azani2018ChEJ} looks extremely attractive since, in this case, there are no dead ends in  the percolation cluster, i.e., the percolation cluster is identical to its geometrical backbone.

In a two-dimensional continuum percolation, the number density is defined as
\begin{equation}\label{eq:n}
n = \frac{N}{L^2},
\end{equation}
where $N$ is the number of objects randomly deposited onto a square substrate of size $L \times L$ with periodic boundary conditions (PBC). The system is said to percolate when there exists a cluster of intersecting objects that spans the square.

When each object has area $a$, the dimensionless quantity
\begin{equation}\label{eq:eta}
\eta = n a,
\end{equation}
is called the filling factor. For zero-width sticks of length $l$, the area is assumed to be $a=l^2$. In the thermodynamic limit $L \to \infty$, the probability that a percolating cluster appears depends only on the filling factor. The percolation threshold for zero-width sticks is $\eta_c^x = 5.637\,285\,8(6)$  while the percolation threshold for discs is $\eta_c^\circ = 1.128\,087\,37(6)$.\cite{Mertens2012PRE} The transparency [see, e.g., Ref.~\onlinecite{Yi2004JAP}] of a film is proportional to the expected fraction of the film not covered by objects
\begin{equation}\label{eq:T}
T= \mathrm{e}^{- n  a}.
\end{equation}

\section{Methods}\label{sec:methods}
\subsection{Sampling}
The samples were prepared exactly as in Ref.~\onlinecite{Azani2018ChEJ}.
The suspension of 0.003 mg/ml of purified silver nanorings in ethanol was prepared and shacken for 1~h using an orbital shaker in order to obtain a uniform suspension. Samples ($5 \times5$~cm) of PET were cut and fixed onto a pre-heated heater at 100~$^\circ$C using high-heat resistance adhesive tape (to prevent the ``coffee ring effect'' during spraying). Spray coating of the nanoring suspension was carried out using a DH-115 SPARMAX airbrush with a nozzle size of 0.35~mm and a 7~ml side feed fluid cup. This spray operates at a pressure of 0.18 to 0.20~MPa. Four transparent conductive films with different sheet resistances were prepared [Table~\ref{tab:samples}]. SEM images were taken using a Hitachi Tabletop microscope (model TM3030) with a magnification range of 15 to 30000~V. The microscope has a pre-centered cartridge filament as an electron gun and a high-sensitivity semiconductor 4-segment BSE detector as its single detection system. This system operates at room temperature in ambient air conditions. The images were processed using the TM3030 software. Optical parameters (haze and transparency) were measured with haze-gard i from BYK, according to illuminants of ASTM-D1003.
 \begin{table}[!htbp]
 \caption{Spray of nanorings on PET substrate. Bare PET: Transparency  91\% and haze 0.6\%.
\label{tab:samples} }
 \begin{tabular}{|c|c|c|c|c|}
 \hline
Sample & Transparency (\%) & Haze & $R_\Box (\Omega/\Box)$ & $n_\text{rel}$ \\
\hline
1 & 84 & 8 & 20 & 3.602 \\
2 & 86 & 5.1 & 45 & 2.543 \\
3 & 87 & 3 & 60 & 2.023 \\
4 & 89 & 1.5 & 350 & 1\\
\hline
 \end{tabular}
 \end{table}

\begin{figure}[!htbp]
  \centering
  \includegraphics[width=\columnwidth]{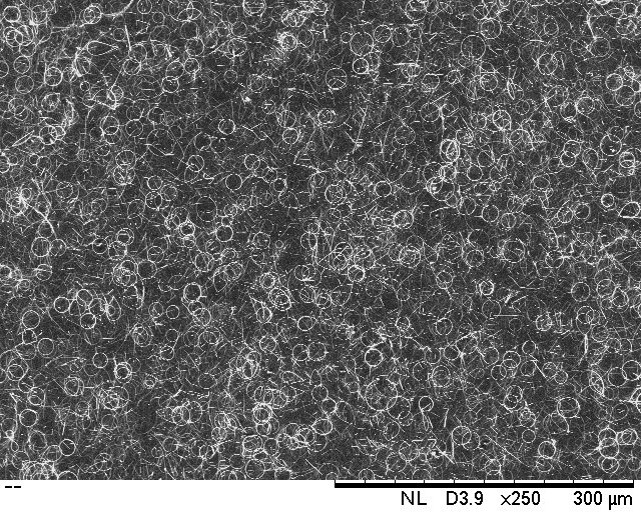}
  \caption{SEM image of Sample 1.\label{fig:sample1}}
\end{figure}

In addition to the rings, a small number of rods were present in the samples [figure~\ref{fig:sample1}]. These rods were ignored both in our theoretical consideration and in the simulation.

\begin{figure}[!htbp]
  \centering
  \includegraphics[width=\columnwidth]{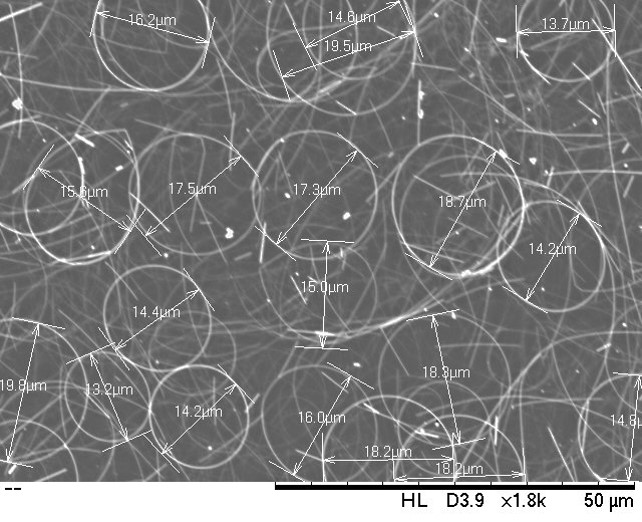}\\%\quad
  \includegraphics[width=\columnwidth]{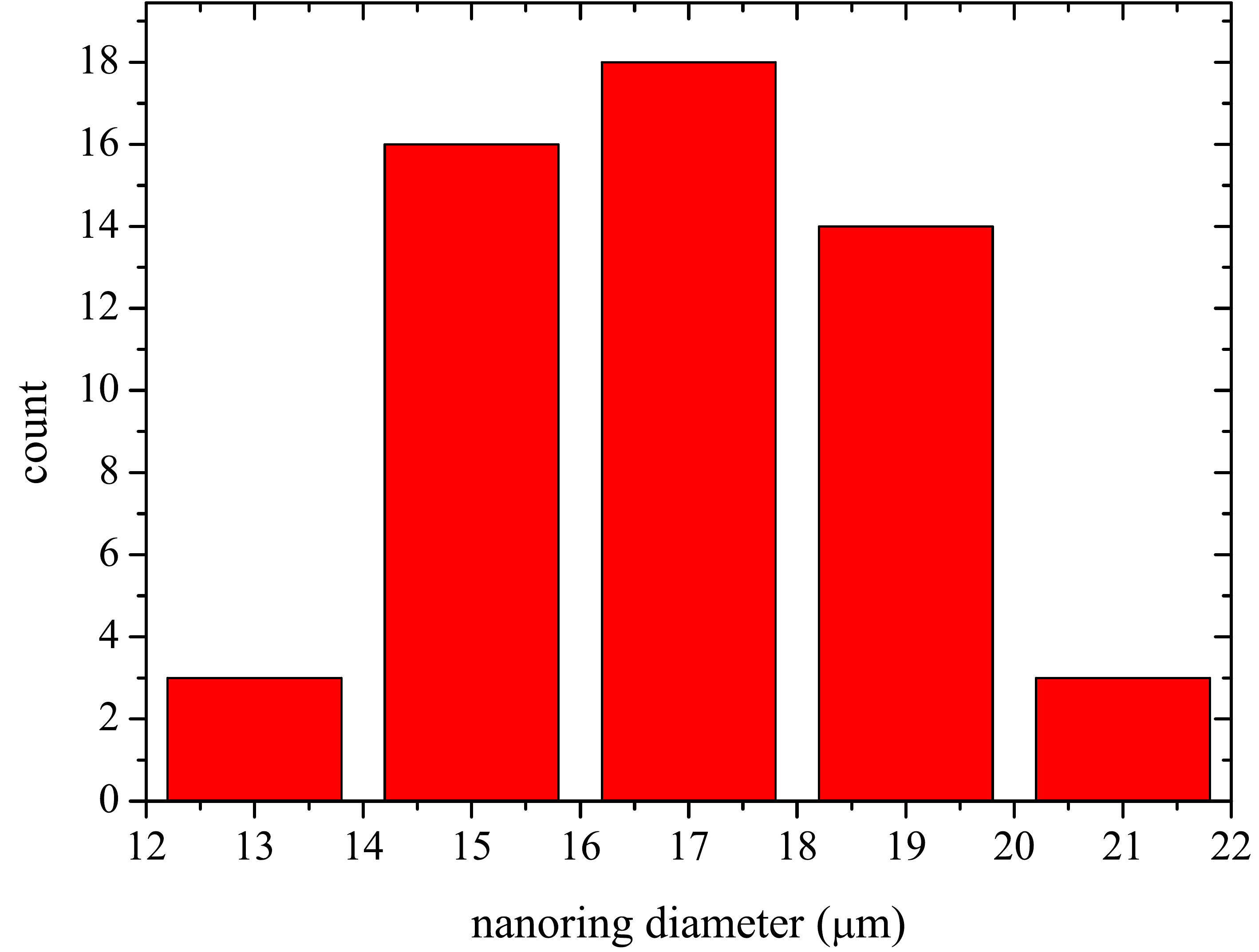}
  \caption{Example of size distribution in Sample 1.\label{fig:sizes}}
\end{figure}

The size dispersity of the rings can be clearly seen [figure~\ref{fig:sizes}].

\subsection{Simulation}
\begin{figure}[!htbp]
 \includegraphics[width=0.9\columnwidth]{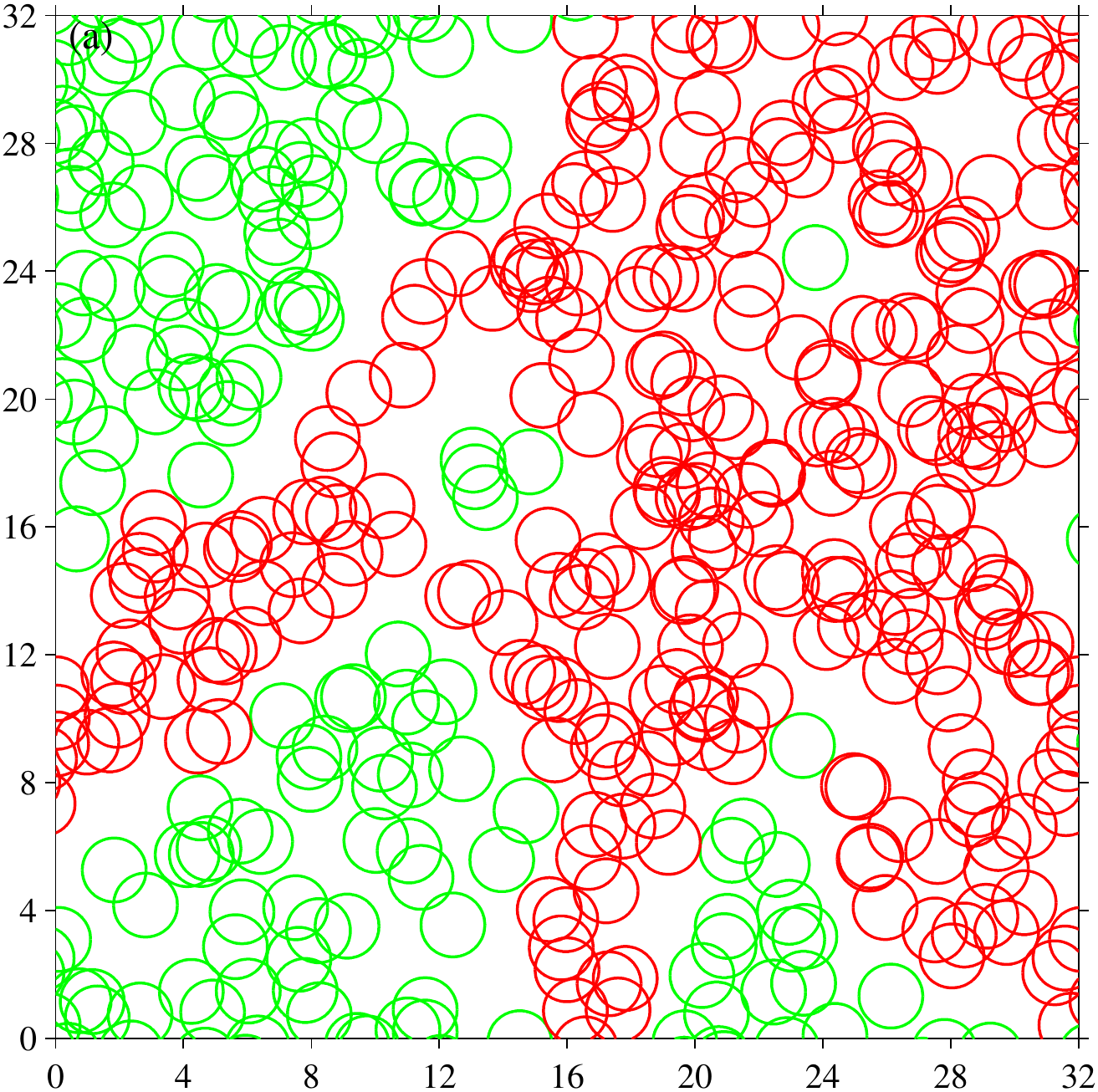}\\%\qquad
 \includegraphics[width=0.9\columnwidth]{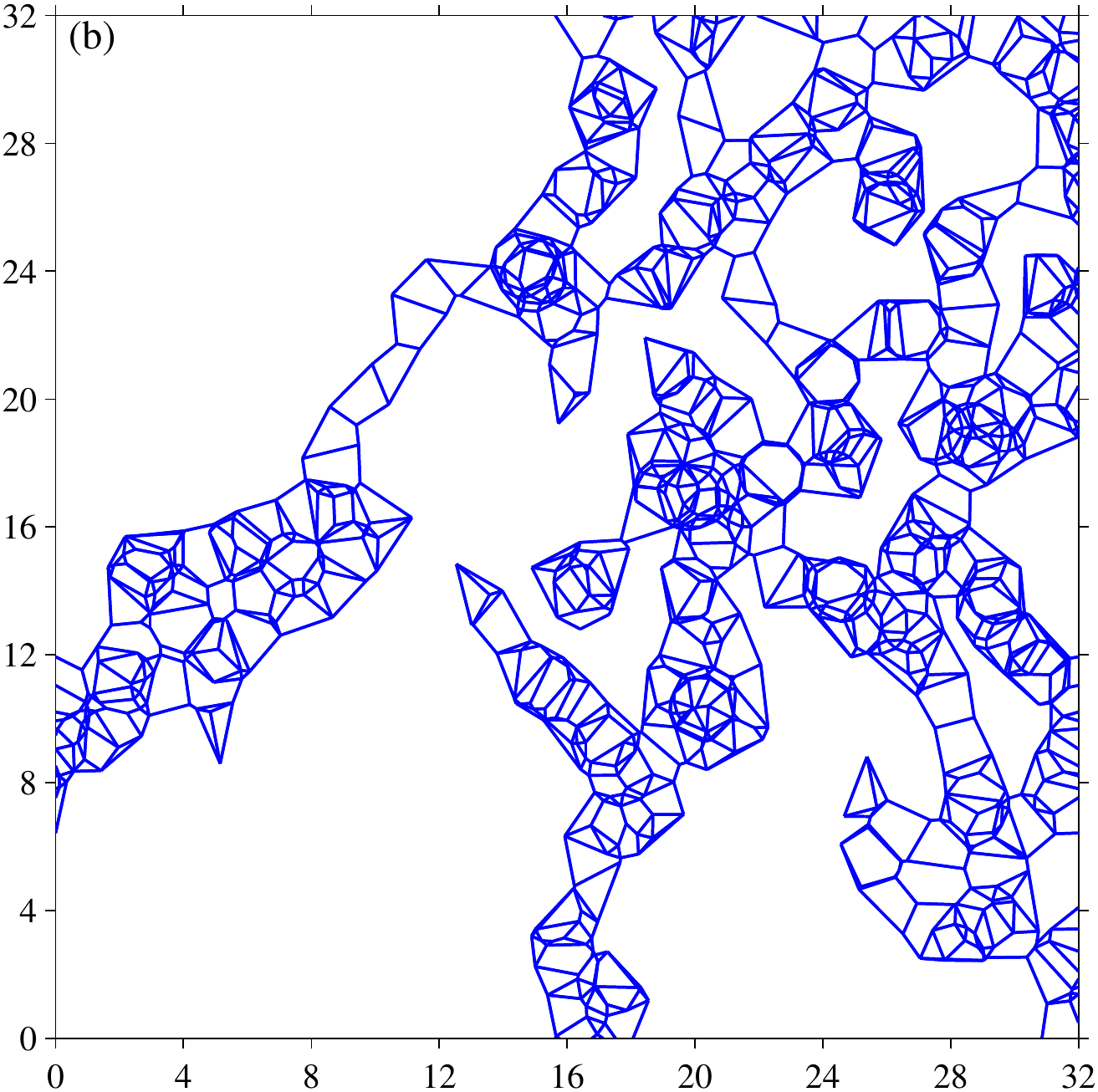}%
 \caption{(a) Example of a simulated system exactly at the percolation threshold. The percolation cluster is shown in red, other clusters are shown in green. The percolation cluster is a weighted multigraph. (b) A simple weighted graph corresponding to the percolation cluster. Multiple edges were collapsed into one edge accompanying by a corresponding recalculation of the weights. Weights of the edges are not indicated for clearer view.\label{fig:depositedrings}}%
 \end{figure}

The rings were randomly deposited onto a substrate of size $L \times L$ with PBC until the desired number density was reached [figure~\ref{fig:depositedrings}]. To detect a spanning cluster, the Union--Find algorithm\cite{Newman2000PRL,Newman2001PRE} modified for continuous systems\cite{Mertens2012PRE,Li2009PRE}  was applied. When a spanning cluster was found, all other clusters were removed since they do not contribute in the electrical conductivity. Then, an adjacency matrix was formed for the spanning cluster, treated as a random weighted  graph. This graph is a multigraph since any two vertices (intersections of two circles) may be connected by more than one edge (arc of a circle). When the length of the $k$-th arc connecting the $i$-th and $j$-th vertices is $l_{ij}^{(k)}$, the weight of the corresponding edge can be taken as $w_{ij}^{(k)} =\left[l_{ij}^{(k)}\right]^{-1}$. To simplify the consideration, each multiple edge was replaced by a single edge with an effective weight $w_{ij} = \sum_k w_{ij}^{(k)}$.  This adjacency matrix was treated as the conductance matrix of a random resistor network (RRN). Then, Kirchhoff's current law was applied to each ring junction, and Ohm's law used for each circuit between two junctions. The resulting set of equations was solved using MATLAB to find the total conductance of the RRN. We ignored any contact resistance in the junctions between rings. The relative number density
\begin{equation}\label{eq:nstar}
  n^\ast = \frac{n}{n_c} -1
\end{equation}
was used, where $n_c$ is the critical number density, i.e., the number density corresponding to the percolation threshold. The electrical conductivities for each particular value of the relative number density were averaged over 10 independent runs and in two mutually perpendicular directions. Two different kinds of systems were simulated, i.e., systems of equal-sized rings and systems of rings with size dispersity. To mimic the real distribution of radii [figure~\ref{fig:sizes}], the beta distribution with probability density function
$$
f(x) = \frac{x^{\alpha-1}(1-x)^{\beta-1}}{\mathrm{B}(\alpha,\beta)},
$$
was used, where $\alpha$ and $\beta$ are the adjusting parameters, and $\mathrm{B}$ is the beta function.

We performed computations for systems of different sizes, viz., $L=16,24,32,48$, and $64$. Since the electrical conductivity depended only weakly on the system size, we present the results for only one particular system size $L=32$.

%\clearpage
\section{Results}\label{sec:results}
\subsection{Simulation}
\begin{figure}[!htbp]
  \centering
  \includegraphics[width=\columnwidth]{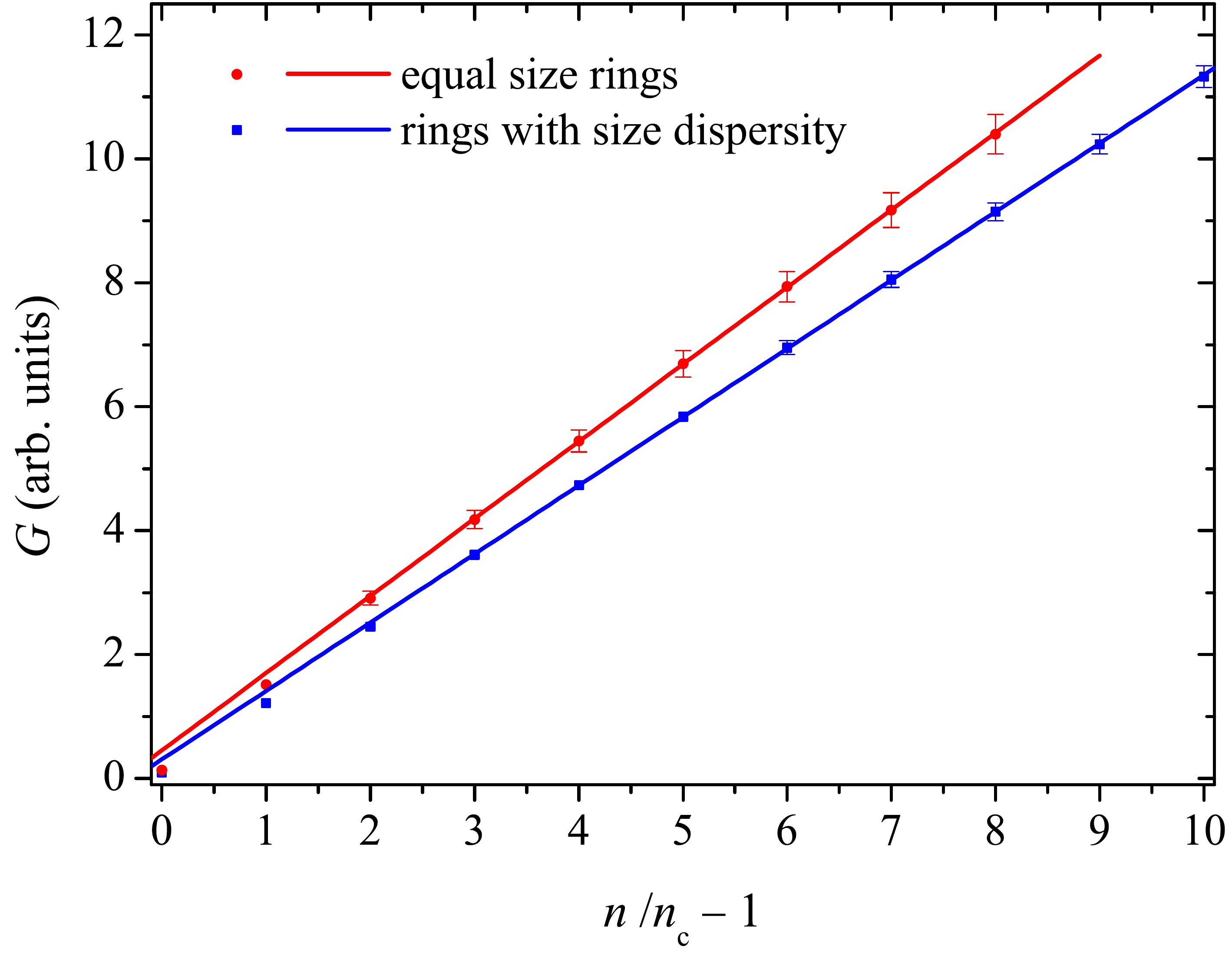}
  \caption{Dependency of the conductance on the relative number of rings per unit area in films with equal-sized rings, and in those with rings of different radii. The error bars correspond to the standard deviation of the mean. When not shown explicitly, they are of the order of the marker size.\label{fig:conductance}}
\end{figure}

Figure~\ref{fig:conductance} demonstrates the dependencies of the electrical conductance on the relative number density for films with equal-sized rings and for films with rings of different sizes. In the both cases, linear growth of the electrical conductance could be observed. A slight deviation from the linear behavior presents only in the vicinity of the percolation threshold. The size dispersity of the rings decreases the percolation threshold, $n_c$. This is the sole reason why the two lines $G(n^\ast)$ have different slopes (while $G(n)$ have the same slopes but different intercepts).

\begin{figure}[!htbp]
  \centering
  \includegraphics[width=\columnwidth]{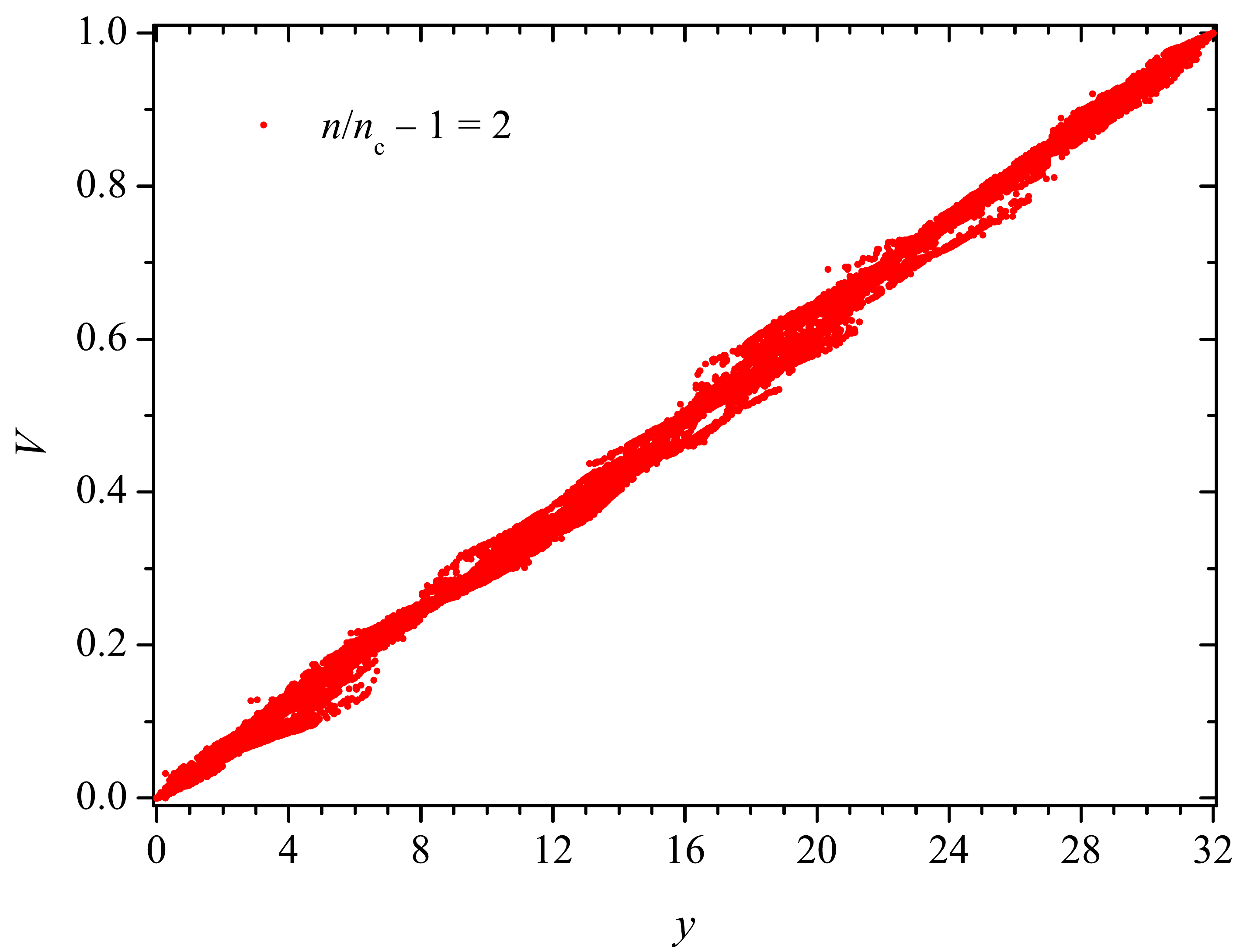}
  \caption{Example of potential distribution in one particular sample with equal-sized rings at $n/n_c -1 = 2$. The potential of each junction in the system is plotted here against its position in the sample.\label{fig:potentials}}
\end{figure}

Figure~\ref{fig:potentials} evidences that a linear change of potential along a sample can be  observed even in not very dense systems. When the number density of rings is three times more than  the critical number density, the potential changes almost linearly.

\begin{figure}
  \centering
  \includegraphics[width=\columnwidth]{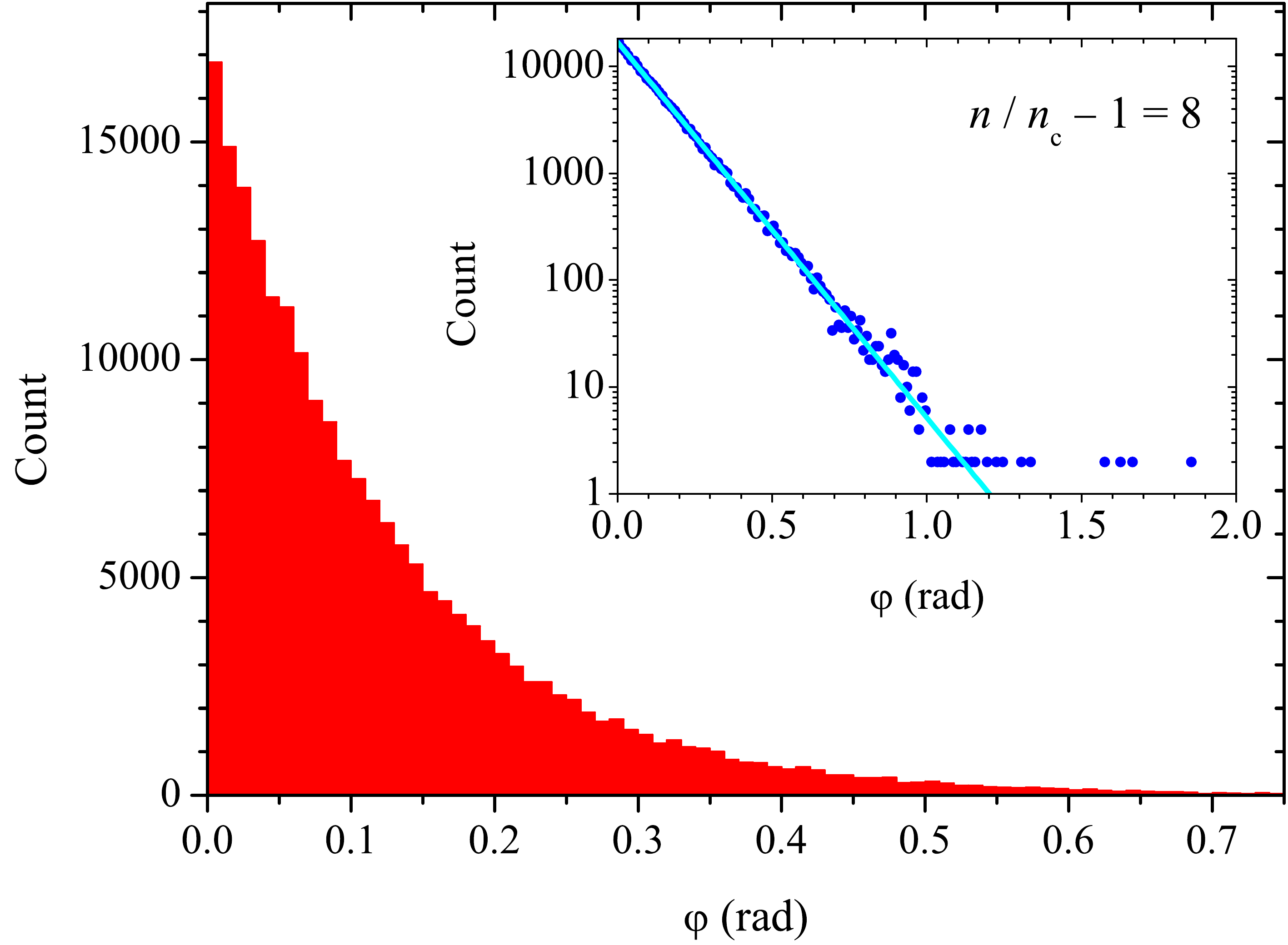}
  \caption{Example of the arc distribution in one particular sample with equal-sized rings at $n/n_c -1 = 8$.\label{fig:arcdistrib}}
\end{figure}

Figure~\ref{fig:arcdistrib} demonstrates  the arc distribution in one particular sample with equal-sized rings at $n/n_c -1 = 8$. The distribution obeys exponential decay. Mainly, only short arcs ($\varphi < 0.5$) are present. Note that $\lambda(0.5)/ l(0.5) \approx 0.99$, hence, for dense systems,  $\lambda(\varphi)/ l(\varphi) \approx 1$.

\subsection{Analytical consideration}
\begin{figure}[!htbp]
  \centering
  \includegraphics[width=0.9\columnwidth]{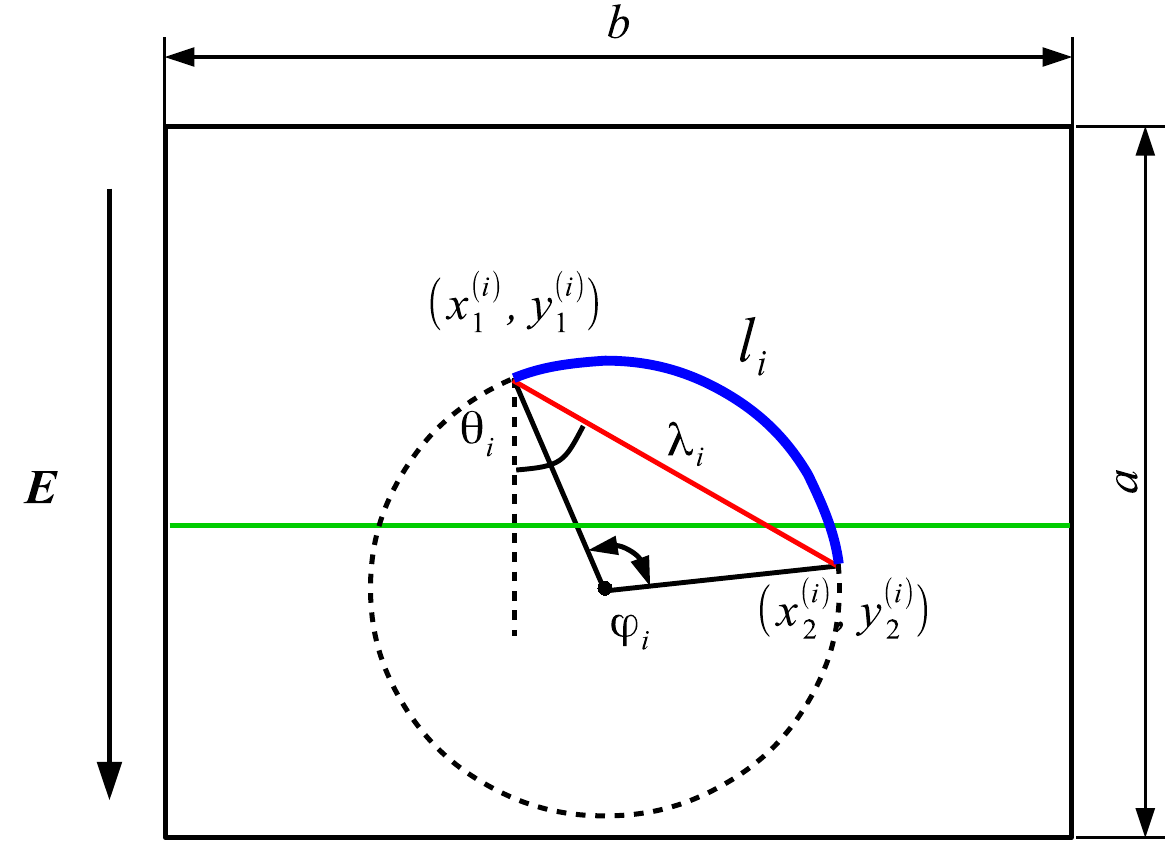}
  \caption{Sketch to illustrate the derivation of sheet resistance.\label{fig:Kumar}}
\end{figure}

A geometrical consideration for dense networks of curved and waved wires\cite{Kumar2016JAP} can directly be adapted for a dense system of rings. Let us consider a rectangular insulating film of size $a \times b$. Conducting rings are randomly deposited onto the film. All rings are assumed to belong to the percolation cluster. This assumption is supported by our simulation, viz., 0.9 of all rings belong to the percolation cluster when at the percolation threshold, while this fraction increases to 0.998 when $n \geqslant 2n^\ast$.  The potential difference $V$ is applied to the opposite borders of the sample [figure~\ref{fig:Kumar}]. The simulations evidence that, when the number density of rings is high, the potential drop along the sample is linear [figure~\ref{fig:potentials}].

We have restricted ourselves to consideration of the simplest case where all the rings have the same radius and the cross-sectional areas of the ring-forming wires are equal and small.
Let us consider an arc between two points of intersection of two rings [figure~\ref{fig:Kumar}]. The potential difference between two junctions is proportional to the length of the projection of the chord between these junctions onto the direction of the electrical field
$$
\Delta V_i = \frac{V \lambda_i \cos \theta_i}{a}.
$$
The electrical conductance of the arc is inversely proportional to its length, $l_i$,
$$
\sigma_i = \sigma \frac{A_{cs}}{l_i},
$$
where $A_{cs}$ is the cross-sectional area of the ring-forming wire, and $\sigma$ is the electrical conductivity of the wire.
The electrical current through the arc is
$$
I_i = \frac{\sigma_i V \lambda_i \cos \theta_i}{a} = \frac{\sigma V A_{cs} \lambda_i \cos \theta_i}{ a l_i}.
$$
The total electrical current through the sample is
$$
I = \frac{\sigma V A_{cs}}{a} \sum_i \frac{\lambda_i}{l_i} \cos \theta_i,
$$
where the summation goes over all those arcs intersecting a horizontal line.
The resistance of the film is
$$
R = \frac{a}{\sigma A_{cs}\sum_i \frac{\lambda_i}{l_i} \cos \theta_i},
$$
while the sheet resistance is
$$
R_\Box = R \frac{b}{a} = \frac{b}{\sigma  A_{cs} \sum_i \frac{\lambda_i}{l_i} \cos \theta_i}.
$$
Arc orientations are expected to be equiprobable and independent on the arc length. This assumption is supported by our simulation with high precision. Moving from summation to integration,
\begin{multline*}
\sum_i \frac{\lambda_i}{l_i} \cos \theta_i \to
\frac{1}{\pi}\int_{-\pi/2}^{\pi/2} \int_{0}^{2\pi} \frac{\lambda(\varphi)}{l(\varphi)} f(\varphi) \, d\varphi \cos \theta \, d\theta =\\ \frac{2}{\pi} \int_{0}^{2\pi} \frac{\lambda(\varphi)}{l(\varphi)} f(\varphi) \, d\varphi,
\end{multline*}
where $\varphi$ is the central angle corresponding to the arc and $f(x)$ is the probability density function (PDF) of the arc lengths. Since the length of the arc is
$l(\varphi) = r \varphi$ and the length of the chord is $\lambda(\varphi) = 2 r \sin \frac{\varphi}{2},$
\begin{multline*}
 \int_{0}^{2\pi} \frac{\lambda(\varphi)}{l(\varphi)}f(\varphi)  \, d\varphi = \int_{0}^{2\pi} \frac{2 \sin \frac{\varphi}{2}}{\varphi} f(\varphi)  \, d\varphi=\\ 2 \int_{0}^{\pi} \frac{\sin z}{z}  f(z) \,dz.
\end{multline*}

When the distance between a ring center and the horizontal line in Figure~\ref{fig:Kumar} does not exceed $r$, this ring is intersected by the line. On average, a horizontal line intersects $2 r b n$ circles. Each circle is intersected twice.

Assuming a dense system, i.e. where all arcs are short, each intersection corresponds to an individual arc, and the total number of intersected arcs is $ 4 r n b$. Hence,
$$
\sum_i \frac{\lambda_i}{l_i} \cos \theta_i \approx \frac{8}{\pi}  r n b .
$$
Therefore, the sheet resistance is
\begin{equation}\label{eq:SheetResistance}
R_\Box = \frac{\pi}{8  \sigma A_{cs} n r}.
\end{equation}
This formula suggests that the reciprocal of sheet resistance, i.e., the sheet conductance, is a linear function of the number of rings per unit area
$$
G = R_\Box^{-1}  = \frac{8  \sigma A_{cs} r n }{\pi}.
$$
Using the relative number density~\eqref{eq:nstar}, the sheet conductance can be written as
\begin{equation}\label{eq:Gsq}
G = R_\Box^{-1}  = \frac{8  \sigma A_{cs} r n_c( 1 + n^\ast) }{\pi}.
\end{equation}
Our simulations were performed for $\sigma A_{cs} r = 1$. For systems of equal-sized rings, $n_c \approx 0.38955$ and  $G \approx 0.45 + 1.246 n^\ast$ [figure~\ref{fig:conductance}]. Our analytical estimation~\eqref{eq:Gsq} suggests $G \approx 0.99 + 0.99 n^\ast$. However, in the vicinity of the percolation threshold ($n \gtrapprox n_c$), the analytical estimation can hardly be treated as reliable since the basic assumptions do not hold.  A difference in slopes is not yet clear since any obvious improvements to the estimations only enlarges this difference.

Note that the expected fraction of the film no covered by rings [Eq.~\eqref{eq:T}], i.e., the transparency of the film, is
  $$
  T  = \exp(- 2 \pi r d n ),
  $$
where $d$ is the diameter of the ring-forming wire. The FoM~\eqref{eq:FoM} in our particular case is
$$
\Phi_\text{TC} = \frac{8 \sigma A_{cs} n r}{\pi}\exp(- 20 \pi r d n ) .
$$
$n_m = (20 \pi r d )^{-1} $ corresponds to the maximal value of the FoM.

\section{Conclusion}\label{sec:concl}
Sheet resistance, transparency, and haze were measured for four samples of transparent conductive films with different numbers of silver nanorings per unit area. Using a Monte Carlo simulation, we studied the electrical conductivity of random resistor networks, obtained by the random deposition of the conductive rings onto an insulating substrate. Our simulation demonstrated a linear dependence of the electrical conductivity on the number of rings per unit area both in the case of equal-sized rings and in the case of rings with size dispersity. While the size dispersity decreased the percolation threshold, the electrical conductance was insensitive to such dispersity.    Analytical estimation of the sheet resistance has been derived for dense systems and the estimation is reasonably consistent with the simulation.

Unfortunately, a direct comparison of experimental and simulation results is hardly possible, since the real samples contain both nanorings and nanowires. Nevertheless, an estimation can be performed. The formula $T = T_0 \exp(- a n)$, where $T_0$ is the bare transparency of the substrate, allows an estimate of the effective concentration of nanoobjects in a sample.
\begin{figure}[!htbp]
  \centering
  \includegraphics[width=\columnwidth]{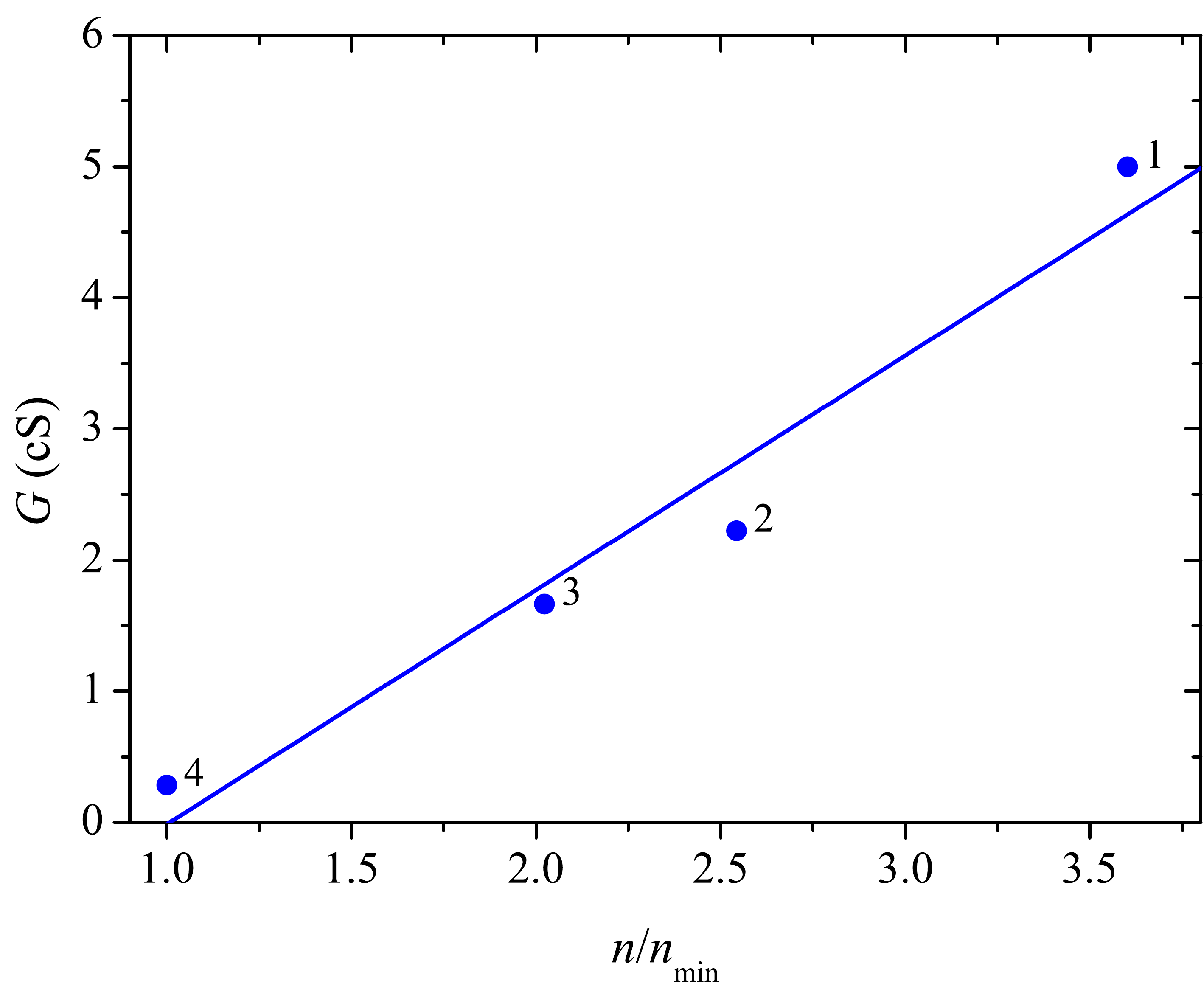}
  \caption{Electrical conductivity of the samples versus the effective number density of nanorings and nanowires. The effective number density is calculated relative to the sample with the highest transparency, i.e., sample 4.\label{fig:Gvsnexp}}
\end{figure}
Figure~\ref{fig:Gvsnexp} demonstrates the dependence of the electrical conductivity of the samples on the effective number density of the nanorings and nanowires. Figure~\ref{fig:Gvsnexp} can be treated as reasonably close to both the simulation and the analytical consideration.  The effective number density is calculated as $\ln (T_i/T_0) / \ln (T_4/T_0)$ [the last column in Table~\ref{tab:samples}].

Theoretical predictions suggest that nanoring-based films are very attractive candidates for the production of transparent electrodes. The linear dependence of the electrical conductance on the number density of conductive nanorings allows for easy prediction of the electrical conductance of any sample. The absence of dead ends in such ring-produced RRNs leads to an efficient use of the conducting elements. However, the manufacturing of pure nanoring-based samples without nanowire admixtures is still a problem requiring resolution..

\begin{acknowledgments}
We would like to acknowledge funding from the Ministry of Science and Higher Education of the Russian Federation, Project No.~3.959.2017/4.6 (Y.Y.T., I.V.V., and A.V.E.), from Intercomet S.L. and from the Ministry of Science, Innovation and Universities of Spain for the sub-program Torres Quevedo and CDTI (M.-R.A. and A.H.).
\end{acknowledgments}

\bibliography{rings}

\end{document}